\documentclass[aps,prb,reprint,superscriptaddress,showpacs]{revtex4-1}

\usepackage{graphicx}
\usepackage{dcolumn}
\usepackage{bm}
\usepackage{amsmath,amssymb}

\begin{document}

\title{Mechanism of uniaxial magnetocrystalline anisotropy in transition metal alloys}

\author{Yohei Kota}
\email[Electronic mail: ]{yohei.kota@aist.go.jp}
\affiliation{Department of Applied Physics, Tohoku University, Sendai 980-8579, Japan}
\affiliation{Spintronics Research Center, AIST, Tsukuba, Ibaraki 305-8568, Japan}
\author{Akimasa Sakuma}
\affiliation{Department of Applied Physics, Tohoku University, Sendai 980-8579, Japan}

\date{\today}

\begin{abstract}
Magnetocrystalline anisotropy in transition metal alloys (FePt, CoPt, FePd, MnAl, MnGa, and FeCo) was studied using first-principles calculations to elucidate its specific mechanism. The tight-binding linear muffin-tin orbital method in the local spin-density approximation was employed to calculate the electronic structure of each compound, and the anisotropy energy was evaluated using the magnetic force theorem and the second-order perturbation theory in terms of spin-orbit interactions. We systematically describe the mechanism of uniaxial magnetocrystalline anisotropy in real materials and present the conditions under which the anisotropy energy can be increased. The large magnetocrystalline anisotropy energy in FePt and CoPt arises from the strong spin-orbit interaction of Pt. In contrast, even though the spin-orbit interaction in MnAl, MnGa, and FeCo is weak, the anisotropy energies of these compounds are comparable to that of FePd,. We found that MnAl, MnGa, and FeCo have an electronic structure that is efficient in inducing the magnetocrystalline anisotropy in terms of the selection rule of spin-orbit interaction.
\end{abstract}

\maketitle

\section{Introduction}

Uniaxial magnetic anisotropy is an essential property of hard magnetic materials, which are widely utilized as permanent magnets and perpendicularly magnetized films. Magnetic anisotropy has several origins, with the main one being magnetocrystalline anisotropy, which is induced by spin-orbit interactions in addition to the anisotropic crystal field.~\cite{chikazumi1997} Since spin-orbit interactions are a relativistic effect on electron motion, a large magnetocrystalline anisotropy has been observed from compounds with heavy elements such as rare-metal and rare-earth elements.

The transition metal systems FePt, CoPt, and FePd, with a tetragonal crystal structure and heavy 4$d$ or 5$d$ elements, show superior performance as hard magnetic materials.~\cite{klemmer1995,farrow1996,gehanno1997,kanazawa2000,okamoto2002,shima2002,shima2004,barmak2005} However, since Pt and Pd are categorized as typical rare-metals, there has been a strong demand for rare-element-free magnets such as MnAl, MnGa, and FeCo in recent years. Both MnAl and MnGa are ferromagnetic and show uniaxial magnetic anisotropy,~\cite{kono1958,koch1960} and a large anisotropy constant has been observed in their film samples.~\cite{cui2011,hosoda2012,nie2013,mizukami2011,mizukami2012,zhu2012} In contrast to MnAl and MnGa, a giant magnetocrystalline anisotropy in FeCo was first predicted theoretically using a first-principles calculation under optimal conditions in terms of Co concentration and tetragonal distortion.~\cite{burkert2004b,neise2011,kota2012c,turek2012} Experimentally, uniaxial magnetic anisotropy was confirmed in artificially strained film samples.~\cite{andersson2006,winkelmann2006,warnicke2007,luo2007,yildiz2009}

First-principles calculations of magnetocrystalline anisotropy in transition metal systems have been carried out for bulk crystals, monolayers, and multilayers.~\cite{daalderop1990,daalderop1991,guo1991,kyuno1992,wang1993,wang1994,daalderop1994,sakuma1994a,sakuma1994b,szunyogh1995,trygg1995,kyuno1996,sakuma1998,halilov1998,ravindran2001} These studies clarified the experimental results of the direction of the magnetic easy axis and the relative magnitude of magnetocrystalline anisotropy energy in real materials on a quantitative level, although the calculation requires a high accuracy because of the small energy scale of the spin-orbit interaction compared with the energy scales of the crystal and exchange fields. Furthermore, a perturbation analysis of the spin-orbit interaction was performed for the magnetocrystalline anisotropy energy of ordered FeNi,~\cite{miura2013} and the physical origin of the perpendicular magnetic anisotropy was quantified.

Magnetocrystalline anisotropy originates from the spin-orbit interaction and the anisotropy energy can be evaluated quantitatively from first-principles calculations; however, a qualitative description of the characteristics in real materials remains an issue. In particular, it is important to clarify the mechanism of magnetocrystalline anisotropy in compounds with no heavy elements. In this paper, we present the characteristics of the uniaxial magnetocrystalline anisotropy of the transition metal alloys FePt, CoPt, FePd, MnAl, MnGa, and FeCo with ordered structures, by using first-principles calculations. We analyze the magnetocrystalline anisotropy energy of these compounds using the magnetic force theorem and the second-order perturbation theory in terms of spin-orbit interactions to elucidate its specific mechanism.

This paper is organized as follows. Details of the calculation method are given in Sec.~2, and the consistency of the perturbation is confirmed in Sec.~3 through a comparison of the magnetocrystalline anisotropy energy calculated using the force theorem and the perturbation theory. In Sec.~4, we discuss the mechanism of the uniaxial magnetocrystalline anisotropy in FePt, CoPt, FePd, MnAl, MnGa, and FeCo. A conclusion is provided in Sec.~5.

\section{Methodology}

\subsection{Electronic structure calculation}

\begin{figure}[t]
\begin{center}
\includegraphics[width=6.0cm,clip]{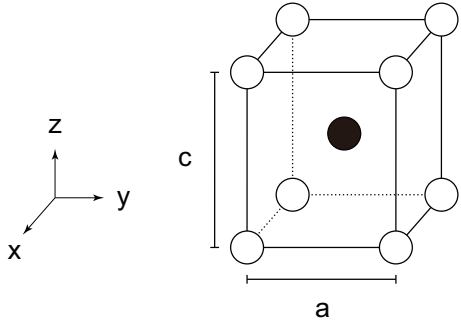}
\caption{Crystal lattice of the distorted $B2$-type structure. The unit cell consists of two sublattices that are located at the corner site (open circles) and the body-center site (closed circles).}
\label{fig1}
\end{center}
\end{figure}

FePt, CoPt, FePd, MnAl, and MnGa ordered alloys have the so-called $L1_0$-type structure, which can be reduced to the tetragonally distorted $B2$-type structure that corresponds to a primitive cell.~\cite{sakuma1994b} Tetragonal FeCo ordered alloy also has the distorted $B2$-type structure. Thus, we used the unit cell of a crystal lattice comprising two atoms, which are located at the corner and body-centered site, as shown in Fig.~\ref{fig1}. The radius of the atomic sphere, which determines the volume of the cell, and the axial ratio $c/a$ were set to the values used in previous studies.~\cite{daalderop1991,sakuma1994a,sakuma1998,kota2012c} We considered the ferromagnetic state of these alloys.

To calculate the electronic structure, we employed the tight-binding linear muffin-tin orbital (TB-LMTO) method in the atomic sphere approximation.~\cite{andersen1975,andersen1984} In this method, the matrix representation of the Kohn-Sham Hamiltonian without the spin-orbit interaction term $\mathcal{H}_0$ is given by
\begin{equation}
\mathcal{H}_0 = C + \sqrt{\Delta} S (1-\gamma S)^{-1} \sqrt{\Delta},
\label{eq1}
\end{equation}
where $C=\{C_{\bm{R}\ell\sigma}\}$, $\Delta=\{\Delta_{\bm{R}\ell\sigma}\}$, and $\gamma=\{\gamma_{\bm{R}\ell\sigma}\}$ are the matrices of the potential parameters in scalar relativistic form, and $S=\{S_{\bm{R}L,\bm{R}'L'}\}$ is the matrix of the canonical structure constant.~\cite{skriver1984,turek1997} Note that the subscripts $\bm{R}$, $L=(\ell,m)$, and $\sigma$ denote the atomic site, orbital, and spin index, respectively. The potential parameters are self-consistently determined in the local spin-density approximation.

\subsection{Evaluation of magnetocrystalline anisotropy energy}

For the evaluation of the magnetocrystalline anisotropy energy, we need to take into account spin-orbit interactions,~\cite{daalderop1994,turek2008}
\begin{equation}
\mathcal{H}_\mathrm{SO} =  \frac{\xi}{2} U(\theta,\phi) (\bm{\ell} \cdot \bm{\sigma}) U^{\dagger}(\theta,\phi),
\label{eq2}
\end{equation}
in addition to $\mathcal{H}_0$. Note that $\xi=\{\xi_{\bm{R}\ell}\}$ is the matrix of the spin-orbit coupling constant, which is estimated by solving the relativistic Dirac equation in a single atomic sphere. Here, $\bm{\ell}$ is the orbital angular momentum operator, $\bm{\sigma}$ is the Pauli matrix, and $U(\theta,\phi)$ is the SU(2) rotation matrix of the spin-1/2 system,~\cite{kubler2009}
\begin{equation}
U(\theta,\phi) = \left(
\begin{array}{cc}
\exp(\frac{i \phi}{2}) \cos(\frac{\theta}{2}) & \exp(-\frac{i \phi}{2}) \sin(\frac{\theta}{2}) \\
-\exp(\frac{i \phi}{2}) \sin(\frac{\theta}{2}) & \exp(-\frac{i \phi}{2}) \cos(\frac{\theta}{2})  \\
\end{array}
\right),
\label{eq3}
\end{equation}
where $\theta$ and $\phi$ are the angles of the direction of magnetization when the unit vector is defined by $\bm{n} = (\sin\theta\cos\phi,\sin\theta\sin\phi,\cos\theta)$. Hereafter, we set $\phi=0$ to focus on the uniaxial anisotropy.

The magnetocrystalline anisotropy energy $\Delta E$ can be evaluated from the energy difference when the magnetization is aligned in the $\theta=0$ and $\theta=\pi/2$ directions. From the magnetic force theorem,~\cite{weinert1985,jansen1999} $\Delta E$ is given by the band energy difference as
\begin{equation}
\Delta E = \sum_{n}^\mathrm{occ} \sum_{\bm{k}} \varepsilon_{\bm{k}n} \vert_{\theta=\frac{\pi}{2}} - \sum_{n}^\mathrm{occ} \sum_{\bm{k}} \varepsilon_{\bm{k}n} \vert_{\theta=0},
\label{eq4}
\end{equation}
where $\varepsilon_{\bm{k}n}$ is the eigenvalue of the Hamiltonian with the spin-orbit interaction $\mathcal{H}_0+\mathcal{H}_\mathrm{SO}$, labeled by the wave vector $\bm{k}$ in the first Brillouin zone and the band index $n$. Using Eq.~(\ref{eq4}) is a common way of evaluating the magnetocrystalline anisotropy energy; however, the physical picture and mechanism is difficult to comprehend.

To address this difficulty, we adopt the expression of the magnetocrystalline anisotropy energy within the second-order perturbation theory in terms of spin-orbit interactions. An expression has been derived in previous works;~\cite{bruno1989,wang1993,laan1998} however, here we present a reformulation of this expression under the same concept: the energy variation $\delta E$ due to spin-orbit interaction is written as
\begin{widetext}
\begin{equation}
\delta E
=
-  \frac{1}{4} \sum^\mathrm{occ}_{n} \sum^\mathrm{unocc}_{n'}  \sum_{\bm{k}} \sum_{\{\bm{R}\}} \sum_{\{L\}} \sum_{\{\sigma\}}
\xi_{\bm{R}\ell} \xi_{\bm{R}'\ell''}
\cdot  \frac{\rho^{\bm{k}n\sigma}_{\bm{R}'L''',\bm{R}L} \rho^{\bm{k}n'\sigma'}_{\bm{R}L',\bm{R}'L''}}{\varepsilon_{\bm{k}n'\sigma'}-\varepsilon_{\bm{k}n\sigma}}
\cdot  \langle L\sigma | U(\bm{\ell} \cdot \bm{\sigma})U^{\dagger} | L'\sigma' \rangle \langle L''\sigma' | U(\bm{\ell} \cdot \bm{\sigma})U^{\dagger} | L'''\sigma \rangle,
\label{eq5}
\end{equation}
and the anisotropy energy $\Delta E = \delta E |_{\theta=\frac{\pi}{2}} - \delta E |_{\theta=0}$ is given by
\begin{subequations}
\label{eq6}
\begin{eqnarray}
\Delta E & = &  \sum_{\{\bm{R}\}}  \sum_{\{L\}}  \sum_{\{\sigma\}}  \Delta E_{\bm{R},\bm{R}'}(L'''L\sigma;L'L''\sigma')
\label{eq6a}  \\
\Delta E_{\bm{R},\bm{R}'}(L'''L\sigma;L'L''\sigma')
& = &
\frac{1}{4}
\sum_{n}^\mathrm{occ}  \sum_{n'}^\mathrm{unocc}  \sum_{\bm{k}}
\xi_{\bm{R}\ell} \xi_{\bm{R}'\ell''}
\cdot  \frac{\rho^{\bm{k}n\sigma}_{\bm{R}'L''',\bm{R}L}  \rho^{\bm{k}n'\sigma'}_{\bm{R}L',\bm{R}'L''}}{{\varepsilon_{\bm{k}n'\sigma'} - \varepsilon_{\bm{k}n\sigma}}}
\cdot \mathcal{C}
\label{eq6b}  \\
\mathcal{C}
& = &  \tau_{\sigma,\sigma'} \cdot  [ \langle L | \ell_z | L' \rangle \langle L'' | \ell_z | L''' \rangle - \langle L | \ell_x | L' \rangle \langle L'' | \ell_x | L''' \rangle ].
\label{eq6c}
\end{eqnarray}
\end{subequations}
\end{widetext}
Note that $\varepsilon_{\bm{k}n\sigma}$ is the eigenvalue of the nonperturbative state, and $\rho^{\bm{k}n\sigma}_{\bm{R}'L',\bm{R}L} = (c^{\bm{k}n\sigma}_{\bm{R}L})^{\ast} c^{\bm{k}n\sigma}_{\bm{R}'L'}$ are products of the expansion coefficients of the eigenstates on an atomic orbital basis, {\it i.e.}, $|\bm{k}n\sigma\rangle = \sum_{\bm{R}L} c^{\bm{k}n\sigma}_{\bm{R}L} | \bm{R}L\sigma\rangle$. We can obtain the coefficient $c^{\bm{k}n\sigma}_{\bm{R}L}$ by solving the secular equation $\det (\varepsilon - \mathcal{H}_0)=0$ directly, since the TB-LMTO is an orthonormal basis in $\bm{R}$, $L$, and $\sigma$. The factor $\tau_{\sigma,\sigma'}$ gives $+1$ for the same-spin ($\sigma=\sigma'$) case and $-1$ for the opposite-spin ($\sigma=-\sigma'$) case. In Eqs.~(\ref{eq5}) and (\ref{eq6}), the summations over terms in curly brackets denote double or quadruple sums, {\it i.e.}, $\{\bm{R}\}=\bm{R},\bm{R}'$, $\{L\}=L,L',L'',L'''$, and $\{\sigma\}=\sigma,\sigma'$. The summation over $n$ ($n'$) is restricted to the occupied (unoccupied) states below (above) the Fermi level $\varepsilon_\mathrm{F}$, which are determined by the condition
\begin{equation}
\int_{-\infty}^{\varepsilon_\mathrm{F}} \sum_{\bm{k}n\sigma} \delta (\varepsilon-\varepsilon_{\bm{k}n\sigma}) d\varepsilon = N_\mathrm{q},
\label{eq7}
\end{equation}
where $N_\mathrm{q}$ is the number of valence electrons of each compound. Approximately $50^3$ $k$-points in the full Brillouin zone were employed for the calculation of the magnetocrystalline anisotropy energy to achieve sufficient accuracy in the following numerical results.

From Eq.~(\ref{eq6}), we are able to conceive a physical picture in which magnetocrystalline anisotropy is attributed to the hybridization between the occupied and unoccupied states through spin-orbit interaction. In particular, such interaction includes the selection rule with respect to the spin and orbital states due to the matrix element of the operators. In addition, the degree of hybridization is dominated by the strength of spin-orbit interaction, $\xi$, and the inverse of the energy difference between the two states, $|\varepsilon^\mathrm{occ}-\varepsilon^\mathrm{unocc}|^{-1}$, on both sides of the Fermi level $\varepsilon_\mathrm{F}$.

\section{Confirmation of Result from Perturbation Theory}

\begin{figure}[b]
\begin{center}
\includegraphics[width=6.0cm,clip]{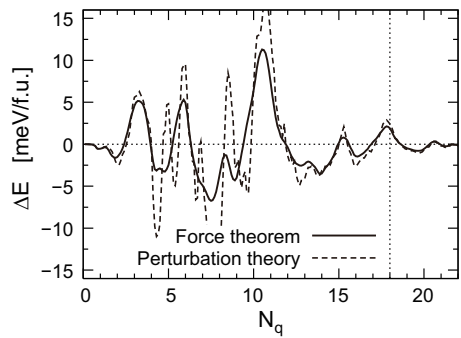}
\caption{Magnetocrystalline anisotropy energy $\Delta E$ in FePt as a function of the number of valence electrons $N_\mathrm{q}$. The solid and dashed lines are the results calculated using the magnetic force theorem in Eq.~(\ref{eq4}) and the second-order perturbation theory in Eq.~(\ref{eq6}), respectively. The actual electron number in FePt is 18.}
\label{fig2}
\end{center}
\end{figure}

\begin{table*}[t]
\caption{Magnetocrystalline anisotropy energies $\Delta E$ in ordered FePt, CoPt, FePd, MnAl, MnGa, and FeCo evaluated using the force theorem (FT) in Eq.~(\ref{eq4}) and perturbation theory (PT) in Eq.~(\ref{eq6}). $c/a$ denotes the axial ratio of tetragonal lattice used in the calculation. Experimental data are also shown for comparison. The number in parentheses shown in the $\Delta E_\mathrm{Exp}$ column of FeCo is $c/a$ in experimental samples.}
\label{table1}
\begin{center}
\begin{tabular}{ccccccc}
\hline
Compound & $c/a$ & $\Delta E_\mathrm{FT}$ [meV/f.u.] & $\Delta E_\mathrm{PT}$ [meV/f.u.] & $\Delta E_\mathrm{FT}$ [MJ/m$^3$] & $\Delta E_\mathrm{PT}$ [MJ/m$^3$]  &  $\Delta E_\mathrm{Exp}$ [MJ/m$^3$]  \\
\hline
FePt  &  1.36  &  1.90 &  2.41  &  11.01  &  14.00  &  $\sim$10 \cite{farrow1996}~, 6.8 \cite{kanazawa2000}~, 5.0 \cite{okamoto2002}~, 4.1 \cite{shima2002}~, 5.5 \cite{barmak2005}  \\
CoPt  &  1.38  &  0.68  &  0.77  &  4.12  &  4.66  &  2.1 \cite{kanazawa2000}~, 3.0 \cite{barmak2005}  \\
FePd  &  1.36  &  0.29  &  0.33  &  1.70  &  1.93  &  1.5 \cite{gehanno1997}~, 2.1 \cite{shima2004}  \\
MnAl  &  1.28  &  0.34  &  0.37  &  1.98  &  2.15  &  1.0 \cite{hosoda2012}~, 1.4 \cite{nie2013}  \\
MnGa  &  1.32  &  0.37  &  0.41  &  2.33  &  2.55  &  1.6 \cite{mizukami2012}~, 2.2 \cite{zhu2012}  \\
FeCo  &  1.15  &  0.23  &  0.20  &  1.61  &  1.39  &   \\
FeCo  &  1.25  &  0.88  &  1.30  &  6.09  &  9.01  &  2.9 (1.18) \cite{andersson2006}~, $>$~0 (1.24) \cite{yildiz2009}  \\
FeCo  &  1.35  &  0.34  &  0.38  &  2.36  & 2.63 &  \\
\hline
\end{tabular}
\end{center}
\end{table*}

We began by confirming that the calculation of the magnetocrystalline anisotropy energy based on the second-order perturbation theory [Eq.~(\ref{eq6})] is consistent with the calculation based on the magnetic force theorem [Eq.~(\ref{eq4})]. Figure~\ref{fig2} shows the $\Delta E$ curve of FePt calculated by both approaches as a function of the number of valence electrons $N_\mathrm{q}$. Note that the observed magnetocrystalline anisotropy energy corresponds to that at $N_\mathrm{q}~=~18$, which is the actual number of valence electrons in FePt. The $N_\mathrm{q}$ dependence was evaluated within the rigid-band scheme using the intrinsic electronic structure of FePt ($N_\mathrm{q}~=~18$) as the self-consistent solution. An investigation of the $N_\mathrm{q}$ dependence is useful for discussing the relationship between the physical quantities and electronic structure as well as the consistency of the calculation.

In the comparison shown in Fig.~\ref{fig2}, the absolute values and trends of the two $\Delta E$ curves are in good agreement with each other. There is very little difference between the results apart from the disagreement in the $5~<~N_\mathrm{q}~<~10$ region where the hypothetical Fermi level is located at the center of the $d$-orbital bands of Pt, which has strong spin-orbit interaction. Table~\ref{table1} shows the magnetocrystalline anisotropy energy calculated using the force theorem in Eq.~(\ref{eq4}) and the perturbation theory in Eq.~(\ref{eq6}) for each compound. In FePt, CoPt, FePd, MnAl, and MnGa, the $\Delta E$ values obtained from the perturbation theory are quantitatively consistent with those obtained from the force theorem, since the magnitude relation in these materials is the same and the numerical deviation between the two results is less than 20\%. Furthermore, the calculated results reproduce the experimental results with respect to the relative magnitude of the magnetocrystalline anisotropy energy in these materials, although most of the experimental results are smaller than the calculation results. One of the reasons for this is that magnetocrystalline anisotropy energy is sensitive to the axial ratio and the degree of order.~\cite{kota2012a,kota2012b} We adopted a constant $c/a$ and a perfectly ordered structure in the calculations; however, the axial ratio and the degree of order depend on the sample preparation conditions in experiments.

In a similar manner, the $\Delta E$ values obtained from the force theorem and the perturbation theory are consistent with each other in FeCo for $c/a~=~1.15$ and~1.35; however, there is a large disagreement between the results in FeCo for $c/a~=~1.25$, for which a giant magnetocrystalline anisotropy has been predicted theoretically.~\cite{burkert2004b,kota2012c,turek2012} The deviation between the results obtained by the two approaches is more than 30\%, even though the strength of the spin-orbit interaction in Fe and Co is one order of magnitude smaller than that in Pt. When we looked at the two $\Delta E$ values as a function of the axial ratio, the disagreement becomes significant as $c/a$ approaches 1.25. This large disagreement is a result of the perturbation failing at approximately $c/a~=~1.25$, {\it i.e.}, the assumption $|\varepsilon^\mathrm{occ}-\varepsilon^\mathrm{unocc}|\gg\xi$ in Eqs.~(\ref{eq5}) and (\ref{eq6}) is inconsistent in some $\bm{k}$-areas, as discussed in the next section.

\section{Uniaxial Magnetocrystalline Anisotropy Mechanism}

We will discuss the characteristics of magnetocrystalline anisotropy in FePt, CoPt, and FePd in Sec.~4.1, MnAl and MnGa in Sec.~4.2, and FeCo in Sec.~4.3. Also we will overview the magnetocrystalline anisotropy mechanism in each compound briefly before presenting the numerical results.

The magnetocrystalline anisotropy energy in FePt and CoPt originates from the heavy Pt atom, whose spin-orbit interaction is much larger than those of the other atoms, and thus, the large coupling constant $\xi$ in Eq.~(\ref{eq6}) results in a larger $\Delta E$. In contrast, MnAl, MnGa, and FeCo for $c/a~=~1.25$ show a large magnetocrystalline anisotropy energy that is comparable to that of FePd in spite of the absence of a heavy element, because the electronic band structure of these compounds is efficient in increasing the magnetocrystalline anisotropy energy. In MnAl and MnGa, $\Delta E$ is dominated by the hybridization between the occupied and unoccupied states that are located near the Fermi level through spin-orbit interaction, specifically, the occupied $d^{\downarrow}_{x^2-y^2}$ and unoccupied $d^{\downarrow}_{xy}$ states, the occupied $d^{\uparrow}_{3z^2-r^2}$ and unoccupied $d^{\downarrow}_{yz}$ states, and the occupied $d^{\uparrow}_{yz}$ and unoccupied $d^{\downarrow}_{3z^2-r^2}$ states. [The notations $\uparrow$ and $\downarrow$ are used to denote the majority- and minority-spin states, respectively, and $L=\{d_{xy}, d_{yz}, d_{xz}, d_{x^2-y^2}, d_{3r^2-z^2} \}$ is the $d$-orbital state ($\ell~=~2$)]. These combinations result in large, positive contributions to $\Delta E$ owing to the coefficient $\mathcal{C}$ that determines the selection rule of spin-orbit interaction in Eq.~(\ref{eq6}). Furthermore, the electronic structure of FeCo for $c/a~=~1.25$ is a special case. The energies of two particular bands that mainly consist of the $d^\downarrow_{xy}$ and $d^\downarrow_{x^2-y^2}$ states on the upper and lower sides of the Fermi level around the $\Gamma$-point are coincidentally close. In this area, $\Delta E$ is significantly high, since the energy difference of these states, $|\varepsilon^\mathrm{occ}-\varepsilon^\mathrm{unocc}|$, is quite small ($\ll \xi$). This implies that the perturbation assumption is no longer valid.

For the numerical calculations, we employed the perturbation theory of the magnetocrystalline anisotropy energy given by Eq.~(\ref{eq6}) for FePt, CoPt, FePd, MnAl, and MnGa. We focused on the contributions of several components such as the atomic site $\bm{R}$, the orbital $L=(\ell,m)$, and the spin $\sigma$. For FeCo, we used the magnetic force theorem given by Eq.~(\ref{eq4}).

\subsection{FePt, CoPt, FePd}

\begin{figure}[t]
\begin{center}
\includegraphics[width=5.9cm,clip]{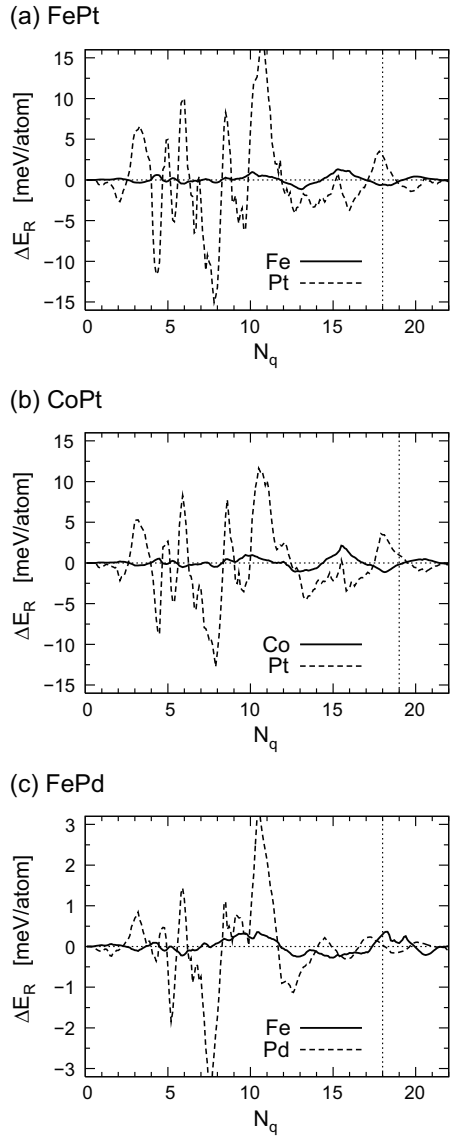}
\caption{Site-decomposed magnetocrystalline anisotropy energy $\Delta E_{\bm{R}}$ as a function of the number of valence electrons $N_\mathrm{q}$ in (a) FePt, (b) CoPt, and (c) FePd. The actual electron number are 18 in FePt and FePd, and 19 in CoPt.}
\label{fig3}
\end{center}
\end{figure}

The $\Delta E$ values of FePt, CoPt, and FePd were decomposed into site $\bm{R}$ contributions because the strength of the spin-orbit interaction varies depending on the atomic species. The spin-orbit coupling constants of Fe, Co, Pd, and Pt are $\xi_{\mathrm{Fe},3d}~=~54$~meV, $\xi_{\mathrm{Co},3d}~=~71$~meV, $\xi_{\mathrm{Pd},4d}~=~189$~meV, and $\xi_{\mathrm{Pt},5d}~=~554$~meV, respectively.~\cite{kota2012b} We defined the expression of the site-decomposed magnetocrystalline anisotropy energy as
\begin{equation}
\Delta E_{\bm{R}}  =  \sum_{\bm{R}'}  \sum_{\{L\}} \sum_{\{\sigma\}}
\Delta E_{\bm{R},\bm{R}'}(L'''L\sigma;L'L''\sigma'),
\notag
\end{equation}
which includes the interference effect with other sites ($\bm{R}$ $\ne$ $\bm{R}'$) via the $\bm{R}'$ summation. Figure~\ref{fig3} shows  the site-decomposed $\Delta E_{\bm{R}}$ values of FePt, CoPt, and FePd as a function of the number of valence electrons $N_\mathrm{q}$. The $\Delta E_{\bm{R}}$ contributions of Pt and Pd are larger than those of Fe and Co with respect to the amplitude because of the stronger spin-orbit interaction in Pt and Pd.

Focusing on the values at the actual electron number ($N_\mathrm{q}~=~18$ for FePt and FePd, and $N_\mathrm{q}~=~19$ for CoPt), the $\Delta E_{\bm{R}}$ contributions of Pt in FePt and CoPt and that of Fe in FePd are predominant in Fig.~\ref{fig3}. We also confirmed that the calculated results of $\Delta E_{\bm{R}}$ at the actual electron number are consistent with those of a previous study.~\cite{solovyev1995} In FePt and CoPt, the Pt-based state is still located close to the Fermi level (the local densities of states in FePt, CoPt, and FePd are shown in Refs.~35 and~41); therefore, the Pt-based state contributes to an increase in the magnetocrystalline anisotropy of FePt and CoPt because of the large spin-orbit interaction of Pt, which is about ten times as large as those of Fe and Co. The roles of the Fe and Co atoms are to induce exchange splitting in the Pt atom via orbital hybridization,~\cite{solovyev1995} even though Pt is nonmagnetic. Consequently, the magnetocrystalline anisotropy mechanism in FePt and CoPt can be described by the synergistic effect between the large spin-orbit interaction in Pt and the large exchange splitting in Fe and Co.

In contrast, the Pd contribution to the magnetocrystalline anisotropy of FePd is not as large, as can be observed from the fact that the $\Delta E_{\bm{R}}$ curve for Pd crosses zero near $N_\mathrm{q}~=~18$. Thus, the magnetocrystalline anisotropy mainly originates from Fe, and the Pd contribution appears to be very small.

\subsection{MnAl, MnGa}

\begin{table*}[t]
\caption{Partial contributions of the magnetocrystalline anisotropy due to the hybridization between the occupied $L\sigma$ and unoccupied $L'\sigma'$ states, $\Delta E^{L\sigma;L'\sigma'}_{\bm{R}}$, in MnAl and MnGa. The unit of energy is meV/Mn~atom. The first group (rows 3--6) and second group (rows 7--12) of the table correspond to the contribution from the nonvanishing matrix elements $\langle L | \ell_z | L' \rangle$ and $\langle L | \ell_x | L' \rangle$, respectively.}
\label{table2}
\begin{center}
\begin{tabular}{cccccccccccc}
\hline
occupied & unoccupied & & \multicolumn{4}{c}{MnAl} & & \multicolumn{4}{c}{MnGa}  \\
\cline{4-7} \cline{9-12}
$L$  &  $L'$  & &  $\Delta E_{\bm{R}}^{L\uparrow;L'\uparrow}$  &  $\Delta E_{\bm{R}}^{L\uparrow;L'\downarrow}$  &  $\Delta E_{\bm{R}}^{L\downarrow;L'\uparrow}$  &  $\Delta E_{\bm{R}}^{L\downarrow;L'\downarrow}$  & &  $\Delta E_{\bm{R}}^{L\uparrow;L'\uparrow}$  &  $\Delta E_{\bm{R}}^{L\uparrow;L'\downarrow}$  &  $\Delta E_{\bm{R}}^{L\downarrow;L'\uparrow}$  &  $\Delta E_{\bm{R}}^{L\downarrow;L'\downarrow}$  \\
\hline
$d_{yz}$          &  $d_{zx}$          & &  0.01 & $-$0.08 & $-$0.01 &  0.07  & &  0.02 & $-$0.07 & $-$0.01 &  0.05  \\
$d_{zx}$          &  $d_{yz}$          & &  0.01 & $-$0.08 & $-$0.01 &  0.07  & &  0.02 & $-$0.07 & $-$0.01 &  0.05  \\
$d_{xy}$          &  $d_{x^2-y^2}$  & &  0.16 & $-$0.22 & $-$0.06 &  0.11  & &  0.14 & $-$0.21 & $-$0.06 &  0.08  \\
$d_{x^2-y^2}$  &  $d_{xy}$          & &  0.02 & $-$0.36 & $-$0.01 &  0.64  & &  0.02 & $-$0.29 & $-$0.01 &  0.47  \\
\hline
$d_{zx}$           &  $d_{xy}$                   & &  $-$0.00 &  0.13 &  0.00 & $-$0.07  & &  $-$0.00 &  0.10 &  0.00 & $-$0.04  \\
$d_{xy}$           &  $d_{zx}$                   & &  $-$0.02 &  0.13 &  0.01 & $-$0.02  & &  $-$0.02 &  0.10 &  0.00 & $-$0.02  \\
$d_{x^2-y^2}$   &  $d_{yz}$                   & &  $-$0.01 &  0.07 &  0.02 & $-$0.12  & &  $-$0.01 &  0.06 &  0.01 & $-$0.09  \\
$d_{yz}$           &  $d_{x^2-y^2}$           & &  $-$0.03 &  0.05 &  0.03 & $-$0.04  & &  $-$0.03 &  0.05 &  0.03 & $-$0.03  \\
$d_{3z^2-r^2}$  &  $d_{yz}$                   & &  $-$0.05 &  0.29 &  0.02 & $-$0.19  & &  $-$0.05 &  0.25 &  0.03 & $-$0.13  \\
$d_{yz}$           &  $d_{3z^2-r^2}$          & &  $-$0.07 &  0.30 &  0.05 & $-$0.21  & &  $-$0.05 &  0.26 &  0.03 & $-$0.12  \\
\hline
\end{tabular}
\end{center}
\end{table*}

\begin{figure}[t]
\begin{center}
\includegraphics[width=6.0cm,clip]{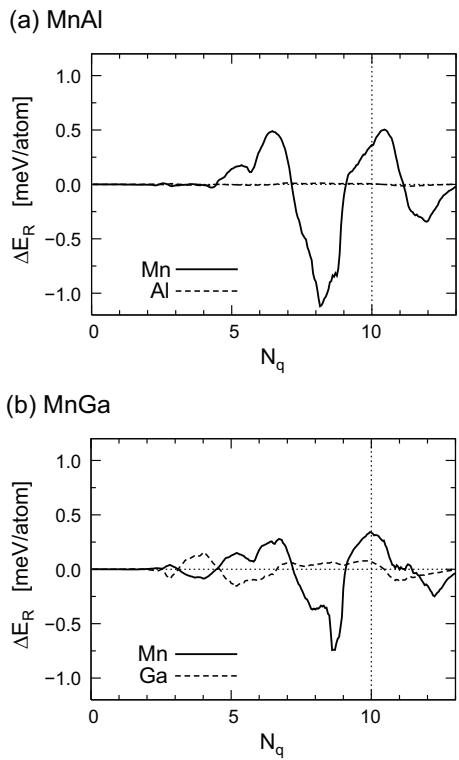}
\caption{Site-decomposed magnetocrystalline anisotropy energy $\Delta E_{\bm{R}}$ as a function of the number of valence electrons $N_\mathrm{q}$ in (a) MnAl and (b) MnGa. The actual electron number in both MnAl and MnGa is 10.}
\label{fig4}
\end{center}
\end{figure}

\begin{figure*}[t]
\begin{center}
\includegraphics[width=12cm,clip]{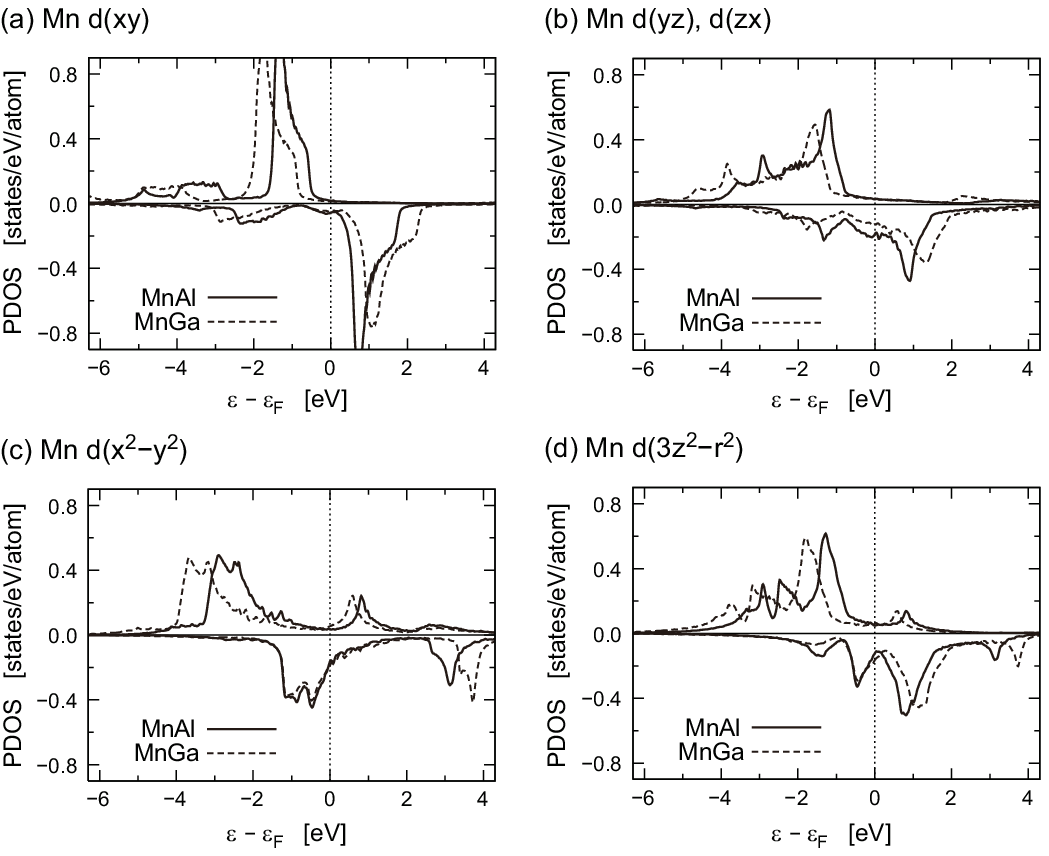}
\caption{Partial densities of states of the Mn $d$-orbitals, (a) $d_{xy}$, (b) $d_{yz}$, $d_{zx}$, (c) $d_{x^2-y^2}$, and (d) $d_{3z^2-r^2}$, in MnAl (solid line) and MnGa (dashed line). Positive and negative values on the vertical axis indicate the densities in the majority- and minority-spin states, respectively.}
\label{fig5}
\end{center}
\end{figure*}

Figure~\ref{fig4} shows the $\Delta E_{\bm{R}}$ values of MnAl and MnGa as a function $N_\mathrm{q}$. The actual electron number of both MnAl and MnGa is $N_\mathrm{q}~=~10$, and the spin-orbit coupling constants of Mn, Al, and Ga are $\xi_{\mathrm{Mn},3d}~=~40$~meV, $\xi_{\mathrm{Al},3p}~=~20$~meV, and $\xi_{\mathrm{Ga},4p}~=~155$~meV, respectively. The $\Delta E_{\bm{k}}$ contribution of Mn is large compared with those of Al and Ga at $N_\mathrm{q}~=~10$, although the Mn and Ga contributions in MnGa are comparable for $N_\mathrm{q} < 6$, for which the Ga $p$-orbital state with a large spin-orbit interaction lies around the hypothetical Fermi level. Therefore, the magnetocrystalline anisotropy of both MnAl and MnGa is dominated by the Mn contribution.

Furthermore, a favorable situation that increases the magnetocrystalline anisotropy of MnAl and MnGa is realized, since there is a large positive $\Delta E_{\bm{R}}$ peak at approximately $N_\mathrm{q}~=~10$ in Fig.~\ref{fig4} despite the small amplitude of the $\Delta E_{\bm{R}}$ curves compared with those of FePt, CoPt, and FePd, as shown in Fig.~\ref{fig3}. For a detailed analysis of the $\Delta E$ values in MnAl and MnGa, we investigated the spin- and orbital-resolved magnetocrystalline anisotropy energy of Mn: $\Delta E^{L\sigma;L'\sigma'}_{\bm{R}} = \Delta E_{\bm{R},\bm{R}}(LL\sigma;L'L'\sigma')$. Hereafter, the $\bm{R}$ subscript denotes the Mn site. The energy $\Delta E^{L\sigma;L'\sigma'}_{\bm{R}}$ corresponds to the partial contribution of the magnetocrystalline anisotropy energy originating from the hybridization between the occupied $L\sigma$ and unoccupied $L'\sigma'$ states through the spin-orbit interaction at the Mn site.

Table~\ref{table2} shows the numerical results of the partial contributions $\Delta E^{L\sigma;L'\sigma'}_{\bm{R}}$ of the $d$-state of Mn in MnAl and MnGa. Here, we adopted real spherical harmonics as a basis for the orbital $L$ components, {\it i.e.}, $d_{xy}$, $d_{yz}$, $d_{zx}$, $d_{x^2-y^2}$, and $d_{3z^2-r^2}$. The $L=L'''$~and~$L'=L''$ contributions, for which the coefficient in Eq.~(\ref{eq6c}) is written as $\mathcal{C}~=~\tau_{\sigma,\sigma'} \cdot [ | \langle L | \ell_z | L' \rangle |^2 - | \langle L | \ell_x | L' \rangle |^2]$, are considered in Table~\ref{table2} with respect to the nonvanishing matrix elements of $\langle L | \ell_z | L' \rangle$ and $\langle L | \ell_x | L' \rangle$.~\cite{laan1998} Because of the coefficient $\mathcal{C}$, the hybridization between same-spin states with a nonvanishing matrix element of $\ell_z$ ($\ell_x$) gives a contribution to the decrease of the energy in the system when the magnetization is aligned in the $z$-direction ($x$-direction). The hybridization between different-spin states gives the complementary contribution. The partial contributions of $L=L''$~and~$L'=L'''$, and the others are omitted because the values are close to zero or quite small compared with those of $L=L'''$~and~$L'=L''$.

In Table~\ref{table2}, the largest $\Delta E_{\bm{R}}^{L\sigma;L'\sigma'}$ is the partial contribution of hybridization between the occupied $d_{xy}^{\downarrow}$ and unoccupied $d_{x^2-y^2}^{\downarrow}$ states through spin-orbit interaction: 0.64 and 0.47 meV/Mn~atom in MnAl and MnGa, respectively. In addition, the second (third) largest contribution with a positive $\Delta E_{\bm{R}}^{L\sigma;L'\sigma'}$ is that of the hybridization between the occupied $d_{yz}^{\uparrow}$ and unoccupied $d_{3z^2-r^2}^{\downarrow}$ states (between the occupied $d_{3z^2-r^2}^{\uparrow}$ and unoccupied $d_{yz}^{\downarrow}$ states). Thus, these contributions increase the uniaxial magnetocrystalline anisotropy of MnAl and MnGa.

As described in Sec.~2, the selection rule of spin-orbit interaction is characterized by $\mathcal{C}$ in Eq.~(\ref{eq6c}) and takes values of $\mathcal{C}~=~\pm 4$, $\pm 3$, and $\pm 1$ when we consider the hybridization between the $d$-orbital ($\ell~=~2$) states.~\cite{laan1998} In particular, the hybridization between the same-spin $d_{xy}$ and $d_{x^2-y^2}$ states and the hybridization between the opposite-spin $d_{yz}$ and $d_{3z^2-r^2}$ states give $\mathcal{C}$ values of $+4$ and $+3$, respectively, which are the largest and second-largest positive $\mathcal{C}$ values for $\ell~=~2$. When these occupied and unoccupied states are located near the Fermi level, the magnetocrystalline anisotropy energy is expected to increase because the strength of the hybridization is proportional to the inverse of the energy difference of the two states, $|\varepsilon^\mathrm{occ}-\varepsilon^\mathrm{unocc}|^{-1}$, as expressed by Eq.~(\ref{eq6b}). These are the basic conditions necessary for increasing $\Delta E$, even though the spin-orbit coupling constant $\xi$ is small.

To confirm the numerical results in Table~\ref{table2} and the corresponding discussion, the partial density of states (PDOS), $\rho_{\bm{R}L\sigma}(\varepsilon)~=~\sum_{\bm{k}n} \rho^{\bm{k}n\sigma}_{\bm{R}L,\bm{R}L} \delta(\varepsilon-\varepsilon_{\bm{k}n\sigma})$, is shown in Fig.~\ref{fig5} for each $d$-orbital component in Mn. The shapes of the PDOS of Mn in MnAl and MnGa are similar, although there is a small difference in the peak position. We find that the PDOSs of the occupied $d_{xy}^{\downarrow}$ and unoccupied $d_{x^2-y^2}^{\downarrow}$ states of both MnAl and MnGa are located near the Fermi level. There is a high probability that these two states will hybridize through spin-orbit interaction owing to the small $|\varepsilon^\mathrm{occ}-\varepsilon^\mathrm{unocc}|$; and therefore, the value of $\Delta E_{\bm{R}}^{L\sigma;L'\sigma'}$ is significantly large, as presented in Table~\ref{table2}. Figure \ref{fig5} also includes the PDOSs of the occupied $d_{yz}^{\uparrow}$ and unoccupied $d_{3z^2-r^2}^{\downarrow}$ states and the occupied $d_{3z^2-r^2}^{\uparrow}$ and unoccupied $d_{yz}^{\downarrow}$ states around the Fermi level, and these states additionally contribute to the total $\Delta E$.

The large positive $\Delta E$ peak at approximately $N_\mathrm{q}~=~10$ in Fig.~\ref{fig4} is mainly attributed to the hybridization between the occupied $d_{xy}^{\downarrow}$ and unoccupied $d_{x^2-y^2}^{\downarrow}$ states, the occupied $d_{yz}^{\uparrow}$ and unoccupied $d_{3z^2-r^2}^{\downarrow}$ states, and the occupied $d_{3z^2-r^2}^{\uparrow}$ and unoccupied $d_{yz}^{\downarrow}$ states through spin-orbit interaction. Therefore, the magnetocrystalline anisotropy energy of MnAl and MnGa, which is comparable to that of FePd, originates from the favorable electronic structure in accordance with the selection rule, although the spin-orbit coupling constant of Mn is small.

\subsection{FeCo}

\begin{figure}[t]
\begin{center}
\includegraphics[width=6.0cm,clip]{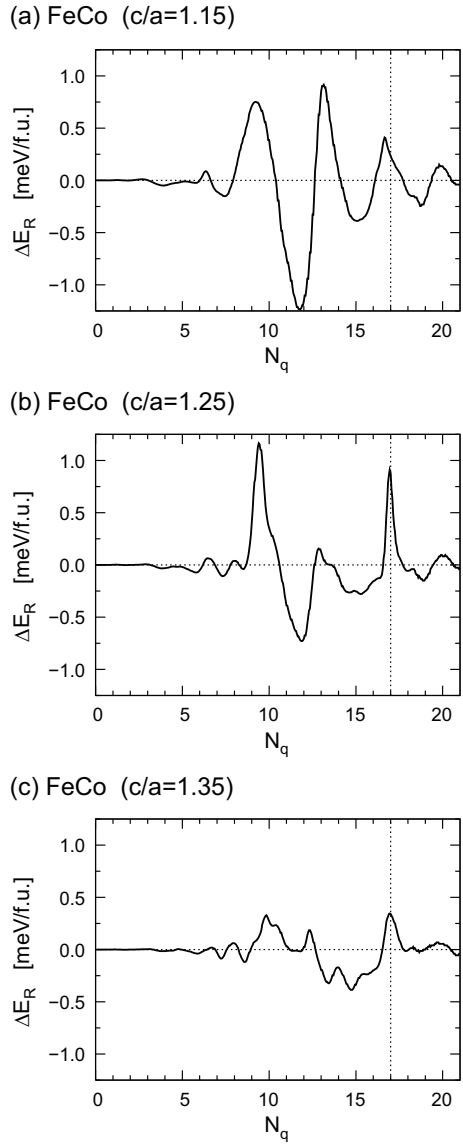}
\caption{Magnetocrystalline anisotropy energy $\Delta E$ as a function of the number of valence electrons $N_\mathrm{q}$ in FeCo for $c/a$ of (a)~1.15, (b)~1.25, and (c)~1.35. The actual electron number is 17.}
\label{fig6}
\end{center}
\end{figure}

\begin{figure*}[t]
\begin{center}
\includegraphics[width=2\columnwidth,clip]{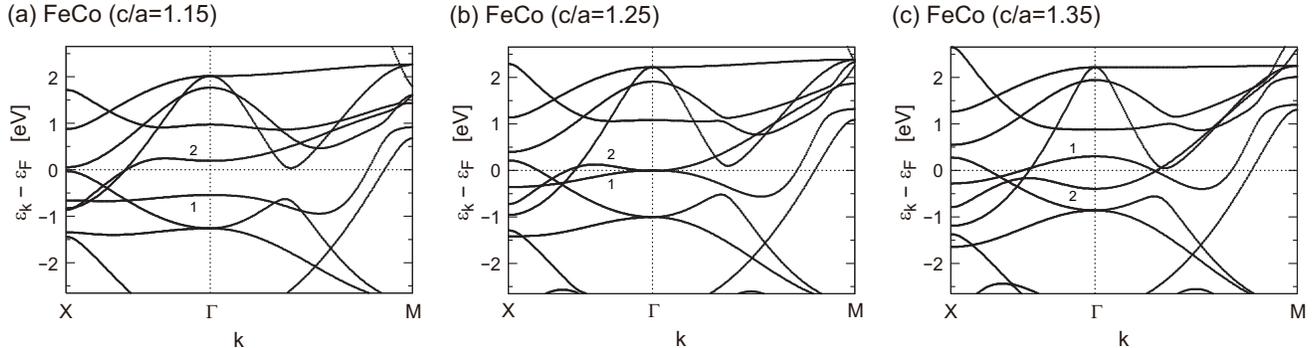}
\caption{Band dispersion $\varepsilon_{\bm{k}}$ in the minority-spin states of FeCo for $c/a$ of (a)~1.15, (b)~1.25, and (c)~1.35 without spin-orbit interactions along the $X$--$\Gamma$--$M$ direction; $\Gamma : (0,0,0)$, $X : (\pi/a,0,0)$, and $M : (\pi/a,\pi/a,0)$. The notations 1 and 2 denote the bands that mainly consist of $d_{xy}$ and $d_{x^2-y^2}$ orbitals, respectively.}
\label{fig7}
\end{center}
\end{figure*}

\begin{figure*}[t]
\begin{center}
\includegraphics[width=2\columnwidth,clip]{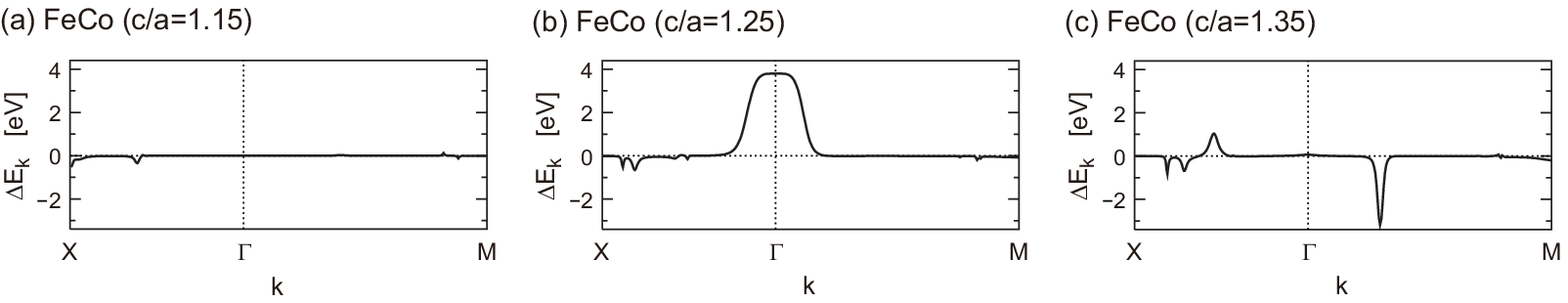}
\caption{$\bm{k}$-resolved magnetocrystalline anisotropy energies $\Delta E_{\bm{k}}$ of FeCo for $c/a$ of (a)~1.15, (b)~1.25, and (c)~1.35 along the $X$--$\Gamma$--$M$ direction; $\Gamma : (0,0,0)$, $X : (\pi/a,0,0)$, and $M : (\pi/a,\pi/a,0)$.}
\label{fig8}
\end{center}
\end{figure*}

A giant magnetocrystalline anisotropy has been predicted from tetragonal Fe$_{1-x}$Co$_x$ disordered alloys under conditions of an axial ratio of $c/a~\sim~1.25$ and Co concentration of $x~\sim~0.5$.~\cite{burkert2004b} Subsequent works have revealed that the magnetocrystalline anisotropy energy increases further in ordered FeCo ($x~=~0.5$) for $c/a \sim~1.25$;~\cite{neise2011,kota2012c,turek2012} thus, we consider ordered FeCo for $c/a~=~1.15$, 1.25, and~1.35 to quantify the giant magnetocrystalline anisotropy. Figure~\ref{fig6} shows the $\Delta E$ values of ordered FeCo for $c/a~=~1.15$, 1.25 and~1.35 as a function of $N_\mathrm{q}$ calculated using the magnetic force theorem given by Eq.~(\ref{eq4}), where the actual electron number of FeCo is $N_\mathrm{q}~=~17$. In FeCo for $c/a~=~1.25$ [Fig.~\ref{fig6}(b)], a sharp $\Delta E$ peak is located just at $N_\mathrm{q}~=~17$; in contrast, $\Delta E$ at $N_\mathrm{q}~=~17$ is not so large for $c/a~=~1.15$ [Fig.~\ref{fig6}(a)] and 1.35 [Fig.~\ref{fig6}(c)].

The origin of the large magnetocrystalline anisotropy in tetragonal Fe-Co alloys has been explained in Ref.~18. The large $\Delta E$ at $N_\mathrm{q}~=~17$ for $c/a~=~1.25$ in Fig.~\ref{fig6}(b) is attributed to the closing of the two particular bands that mainly consist of the $d_{xy}^{\downarrow}$ and $d_{x^2-y^2}^{\downarrow}$ states near the Fermi level around the $\Gamma$-point.

In a body-centered cubic crystal ($c/a~=~1$), the energy bands that mainly consist of the $t_{2g}$ ($d_{xy}$, $d_{yz}$, $d_{zx}$) orbital states are triply degenerate, and the energy bands that mainly consist of the $e_g$ ($d_{x^2-y^2}$, $d_{3z^2-r^2}$) orbital states are doubly degenerate at the $\Gamma$-point. The $t_{2g}$- and $e_g$-based states are located below and above the Fermi level, respectively, in the minority-spin states of Fe-Co alloys. In contrast, in a body-centered tetragonal crystal ($c/a~>~1$), the triple degeneracy in the $t_{2g}$ states and the double degeneracy in the $e_g$ states are resolved, and the energy level of the $d_{xy}$-based ($d_{x^2-y^2}$-based) states shifts upward (downward) with increasing $c/a$.~\cite{burkert2004b}

Figure~\ref{fig7} shows the band dispersion $\varepsilon_{\bm{k}}$ of FeCo calculated without the spin-orbit interaction in the minority-spin states. For $c/a~=~1.15$ [Fig.~\ref{fig7}(a)], there are two bands that mainly consist of the $d_{xy}^{\downarrow}$ and $d_{x^2-y^2}^{\downarrow}$ states (represented by 1 and 2, respectively), which are located on the lower and upper sides of the Fermi level around the $\Gamma$-point. Moreover, when $c/a$ increases, the $d^{\downarrow}_{xy}$- and $d^{\downarrow}_{x^2-y^2}$-based bands move up and down, respectively. For $c/a~=~1.25$ [Fig.~\ref{fig7}(b)], the energy levels of these two bands move closer to each other, and the Fermi level of FeCo ($N_\mathrm{q}~=~17$) is located just at the intermediate energy between the $d^{\downarrow}_{xy}$- and $d^{\downarrow}_{x^2-y^2}$-based bands around the $\Gamma$-point. This situation is very favorable in terms of the conditions for enhancing the uniaxial magnetocrystalline anisotropy because the $d^{\downarrow}_{xy}$ and $d^{\downarrow}_{x^2-y^2}$ states satisfy selection rule of the spin-orbit interaction. For $c/a~=~1.35$ [Fig.~\ref{fig7}(c)], the energy levels of the $d^{\downarrow}_{xy}$- and $d^{\downarrow}_{x^2-y^2}$-based band are reversed.

For the numerical analysis of the magnetocrystalline anisotropy energy of tetragonal FeCo, we looked at the $\bm{k}$-resolved magnetocrystalline anisotropy energy:
\begin{equation}
\Delta E_{\bm{k}} =  \sum_{n}^\mathrm{occ} \varepsilon_{\bm{k}n}
\vert_{\theta=\frac{\pi}{2}} - \sum_{n}^\mathrm{occ}
\varepsilon_{\bm{k}n} \vert_{\theta=0}.
\notag
\end{equation}
Figure~\ref{fig8} shows the $\Delta E_{\bm{k}}$ values of FeCo for $c/a~=~1.15$, 1.25, and~1.35 along the $X$--$\Gamma$--$M$ direction. The $\Delta E_{\bm{k}}$ values around the $\Gamma$-point for $c/a~=~1.25$ [Fig.~\ref{fig8}(b)] are extraordinary large in comparison with those in the other $\bm{k}$-regions and those for $c/a~=~1.15$ [Fig.~\ref{fig8}(a)] and~1.35 [Fig.~\ref{fig8}(c)]. As shown in Fig.~\ref{fig8}(b), the $d^{\downarrow}_{xy}$- and $d^{\downarrow}_{x^2-y^2}$-based bands overlap around the $\Gamma$-point near the Fermi level; therefore, the large magnetocrystalline anisotropy is induced by the hybridization between these states via spin-orbit interaction. In this region, the difference between the energies of the two bands $|\varepsilon^\mathrm{occ}-\varepsilon^\mathrm{unocc}|$ is significantly smaller than the energy scale of the spin-orbit coupling constant $\xi$; thus, the perturbation assumption is no longer valid.

The large $\Delta E$ of FeCo for $c/a~=~1.25$ mostly originates from the $\Delta E_{\bm{k}}$ contribution around the $\Gamma$-point, even though this large $\Delta E_{\bm{k}}$ is enhanced in the small $\bm{k}$-region in the first Brillouin zone. Therefore, the sharp peak at $N_\mathrm{q}~=~17$ appears in FeCo for $c/a~=~1.25$, as shown in Fig.~\ref{fig6}(b). The $\Delta E_{\bm{k}}$ for $c/a~=~1.35$ in Fig.~\ref{fig8}(c) also increases around $(0.54 \ \pi/a,0,0)$ and $(0.29 \ \pi/a,0.29 \ \pi/a,0)$ because some bands cross the Fermi level. However, the range of large $\Delta E_{\bm{k}}$ values is narrow, and the contribution to the total $\Delta E$ is small.

Consequently, the mechanism of the giant magnetocrystalline anisotropy in ordered FeCo is the closing of the energy level between the two bands that mainly consist of the $d^{\downarrow}_{xy}$ and $d^{\downarrow}_{x^2-y^2}$ states near the Fermi level around the $\Gamma$-point for $c/a~=~1.25$. This situation is distinguished from the special cases of MnAl and MnGa; however, the obtained results imply that this particular band structure in FeCo for $c/a~=~1.25$ is strictly dependent on the axial ratio and position of the Fermi level, {\it i.e.}, the number of valence electrons. In addition, the magnetocrystalline anisotropy in FeCo for $c/a~=~1.25$ is strongly influenced by the finite lifetime of electron scattering caused by chemical disorder, according to recent studies using the coherent potential approximation.~\cite{kota2012c,turek2012} If the electron lifetime $\tau$ satisfies the condition $\hbar/2\tau \gg |\varepsilon^\mathrm{occ}-\varepsilon^\mathrm{unocc}|$, then $\Delta E$ decreases considerably as a result of the Bloch spectral function being smeared.~\cite{kota2013}

\section{Conclusions}

We have examined the uniaxial magnetocrystalline anisotropy in FePt, CoPt, FePd, MnAl, MnGa, and FeCo and characterized thespecific mechanisms using first-principles calculations. In our evaluation of the magnetocrystalline anisotropy energy, the numerical results obtained from the second-order perturbation theory in terms of spin-orbit interactions were in quantitative agreement with those obtained from the force theorem as long as the perturbation assumption was valid. We elucidated the mechanism systematically and presented the conditions necessary for increasing the uniaxial magnetocrystalline anisotropy in real materials.

The magnetocrystalline anisotropy energy of FePt and CoPt was shown to originate from Pt, which has a strong spin-orbit interaction. In contrast, a large magnetocrystalline anisotropy compared with that of FePd was observed in MnAl, MnGa, and FeCo, even though the spin-orbit interaction is weak. The mechanism of the uniaxial anisotropy in MnAl and MnGa was described by the electronic structure that the occupied $d_{xy}^{\downarrow}$ and unoccupied $d_{x^2-y^2}^{\downarrow}$ states, the occupied $d_{yz}^{\uparrow}$ and unoccupied $d_{3z^2-r^2}^{\downarrow}$ states, and the occupied $d_{3z^2-r^2}^{\uparrow}$ and unoccupied $d_{yz}^{\downarrow}$ states are located near the Fermi level. This situation is efficient in inducing the uniaxial magnetocrystalline anisotropy in terms of the selection rule for the hybridization of these states through spin-orbit interaction. Furthermore, the electronic structure of FeCo for $c/a~=~1.25$ is a special case of the electronic structure in MnAl and MnGa. The mechanism of the uniaxial anisotropy in FeCo for $c/a~=~1.25$ involves a decrease in the energy difference of the two bands based on the $d_{xy}^{\downarrow}$ and $d_{x^2-y^2}^{\downarrow}$ states on both sides of the Fermi level around the $\Gamma$-point. These results imply that the magnetocrystalline anisotropy energy can be increased even though the spin-orbit interaction in the system is weak, if the electronic band structure satisfies the conditions for the selection rule of spin-orbit interaction.


\begin{acknowledgments}
The authors are grateful to Emeritus Professor Takeshi~Watanabe for encouragement during their study. One of the authors~(Y.K.) was supported by a Grant-in-Aid for JSPS~Fellows.
\end{acknowledgments}

\bibliography{references}

\end{document}